\documentclass[twocolumn,prb,aps,showpacs]{revtex4}

\usepackage{dcolumn}
\usepackage{amsmath}
\usepackage{epsfig}
\newlength{\figwidth}
\preprint{DRAFT}

\begin{document}


\title{Molecular orbital excitations in cuprates}
\author{Young-June Kim}
\author{J. P. Hill}
\author{G. D. Gu}
\affiliation{Department of Physics, Brookhaven National Laboratory,
Upton, New York 11973}
\author{F. C. Chou}
\affiliation{Center for Materials Science and Engineering,
Massacusetts Institute of Technology, Cambridge, Massachusetts
02139}
\author{S. Wakimoto}
\author{R. J. Birgeneau}
\affiliation{Department of Physics, University of Toronto,
Toronto, Ontario M5S 1A7, Canada}
\author{Seiki Komiya}
\author{Yoichi Ando}
\affiliation{Central Research Institute of Electric Power Industry, Komae, Tokyo 201-8511, 
Japan}
\author{N. Motoyama}
\author{K. M. Kojima}
\author{S. Uchida}
\affiliation{Graduate School of Frontier Sciences, University of Tokyo, Bunkyo, Tokyo 113-8656, Japan}
\author{D. Casa}
\author{T. Gog}
\affiliation{CMC-CAT, Advanced Photon Source, Argonne National
Laboratory, Argonne, Illinois 60439}

\date{\today}

\begin{abstract}

We report resonant inelastic x-ray scattering studies of electronic
excitations in a wide variety of cuprate compounds.  Specifically, we focus
on the charge-transfer type excitation of an electron from a bonding
molecular orbital to an antibonding molecular orbital in a copper oxygen
plaquette.  Both the excitation energy and the amount of dispersion are
found to increase significantly as the copper oxygen bond-length is
reduced. We also find that the estimated bond-length dependence of the
hopping integral $t_{pd}$ is much stronger than that expected from
tight-binding theory.

\end{abstract}


\maketitle

\section{introduction}

One of the most important characteristics of the electronic structure
of the cuprates is the strong hybridization between the Cu
3$d_{x^2-y^2}$ level and the O 2$p_\sigma$ level, where $p_\sigma$
denotes the $p_x$ or $p_y$ orbitals pointing towards the Cu ions.
Because of this hybridization, the Cu-O bond has a strong covalent
character and a large energy splitting exists between the bonding
($\sigma$) and antibonding ($\sigma^\ast$) molecular orbitals. In the
ionic limit without hybridization, this energy splitting corresponds to
the energy difference between the atomic $d_{x^2-y^2}$ and $p_\sigma$
orbitals, which is $\Delta_0 \sim 3.5$ eV. \cite{Hybersten89} As the
$p-d$ hybridization becomes larger, the energy splitting
($\Delta_{\sigma\sigma^\ast}$) between the two molecular orbitals
increases, reflecting the increasingly covalent nature of the Cu-O
bonding. Thus, $\Delta_{\sigma\sigma^\ast}$ is a direct measure of the
Cu-O hybridization, and could serve as an independent route to
determine the value of the hopping matrix element, $t_{pd}$, since it
is believed that $\Delta_{\sigma\sigma^\ast}$ is directly related to
$t_{pd}$. \cite{vanVeenendaal93} Although there has been no systematic
theoretical study of $\Delta_{\sigma\sigma^\ast}$, the values obtained
from first-principles calculations range widely from $\sim 4$ eV to
$\sim 9$~eV. \cite{Mattheiss87,McMahan88,Martin93,Husser00} For
example, Martin and Hay \cite{Martin93} have carried out an {\it ab
initio} quantum chemistry calculation of a cluster of copper oxygen
octahedra (CuO$_6^{10-}$) in $\rm La_{1.85}Sr_{0.15}CuO_4$, and
obtained $\Delta_{\sigma\sigma^\ast}\sim 9$ eV, while in a recent
density-functional calculation of a similar cluster, H\"usser and
coworkers reported a value of $\Delta_{\sigma\sigma^\ast}\sim 5.8$~eV.
\cite{Husser00} Experimentally, determining this quantity has been very
difficult, since in this energy range transitions involving the La $4f$
bands dominate the spectral features of optical spectroscopy.
\cite{Uchida91}

In this paper, we report a systematic experimental study of
$\Delta_{\sigma\sigma^\ast}$, the excitation energy from bonding to
antibonding molecular orbitals, using the recently developed resonant
inelastic x-ray scattering (RIXS) technique. \cite{Kao96,Hill98} RIXS
is ideally suited for this study, since it provides element-specific
and momentum-dependent information for electronic excitations.
\cite{Kotani01} By tuning the incident photon energy to the Cu-K
absorption edge, one can gain information concerning excitations
localized around the Cu sites without suffering from problems due to
the La bands. 
Furthermore, the momentum-resolving capability of
RIXS provides additional information: the dispersion of such molecular
orbital (MO) excitations. In this work, we find that  
$\Delta_{\sigma\sigma^\ast}$ exhibits a strong, systematic
dependence on the Cu-O bond length ($d_{\rm Cu-O}$), increasing as
$d_{\rm Cu-O}$ is decreased. In addition, for materials with a small
$d_{\rm Cu-O}$ and correspondingly large $\Delta_{\sigma\sigma^\ast}$,
a relatively large dispersion of the MO excitation is observed. We
discuss the implication of these observations for understanding the
electronic structure of the cuprates.

\begin{table*}
\caption
{The copper oxide samples are listed along with the Cu-O bond lengths 
taken from the references. The top eight
materials possess perfect square copper oxygen plaquettes, while the bottom 
three have distorted square plaquettes. Also listed are 
the polarization and the energy of the incident photon 
in RIXS measurements.} 
\label{table1} 
\begin{ruledtabular} \begin{tabular}{cccccccc}
Label & Sample & Crystal\footnotemark[1] & $d_{\rm Cu-O}$ (\AA) & Ref. 
($d_{\rm Cu-O}$) & 
Polarization & $E_i$ 
(eV) & Ref. (RIXS) \\
\hline
2122& $\rm Sr_2CuO_2Cl_2$& M & 1.9858 &  \onlinecite{Miller90} & 
$\boldsymbol{\epsilon}~\perp~{\bf z}$ & 9001 & 
This 
work \\
Nd & $\rm Nd_2CuO_4$&  & 1.9705 & \onlinecite{Gopalakrishnan89} & 
$\boldsymbol{\epsilon}~\perp~{\bf z}$ & 8990 
& 
\onlinecite{Hill99} \\
Ca & $\rm Ca_2CuO_2Cl_2$& & 1.9344 & \onlinecite{Argyriou95} & 
$\boldsymbol{\epsilon}~\parallel~\sim{\bf z}$ 
\footnotemark[2]
& 8996 & \onlinecite{Hasan00} \\
2342 & $\rm Sr_2Cu_3O_4Cl_2$& M & 1.929 & \onlinecite{2342-PRB} & 
$\boldsymbol{\epsilon}~\perp~{\bf z}$ & 8998 
& 
This work \\
LCCO & $\rm La_{1.9}Ca_{1.1}Cu_2O_6$& B & 1.913 & \onlinecite{Izumi89} & 
$\boldsymbol{\epsilon}~\parallel~{\bf 
z}$ & 
8999 & This work \\
LCO & $\rm La_2CuO_4$& T & 1.904 & \onlinecite{Radaelli94} &  
$\boldsymbol{\epsilon}~\parallel~{\bf z}$ & 
8997 & This work \\
LSCO5 & $\rm La_{1.95}Sr_{0.05}CuO_4$& C & 1.898 & \onlinecite{Radaelli94} 
&  
$\boldsymbol{\epsilon}~\parallel~{\bf z}$ & 
8997 & This work \\
LSCO17 & $\rm La_{1.83}Sr_{0.17}CuO_4$& C & 1.885 & 
\onlinecite{Radaelli94} & 
 $\boldsymbol{\epsilon}~\parallel~{\bf z}$ & 8997 & This 
work\footnotemark[3] \\
Li & $\rm Li_2CuO_2$&  & 1.9577 & \onlinecite{Sapina90} & 
$\boldsymbol{\epsilon}~\perp~{\bf z}$ & 8997 & 
\onlinecite{Li-PRB} \\
CGO & $\rm CuGeO_3$& & 1.9326 & \onlinecite{Braden96} & 
$\boldsymbol{\epsilon}~\parallel~{\bf x}+{\bf z}$ & 
8990 & \onlinecite{Zimmermann02} \\
112 & $\rm SrCuO_2$& U & 1.910/1.930/1.961 \footnotemark[4]
& \onlinecite{Matsushita94} & 
$\boldsymbol{\epsilon}~\parallel~{\bf z}$ & 
8996 & This work \\
\end{tabular}
\end{ruledtabular}
\footnotetext[1]{The crystals studied in this work were provided by 
various groups, which are denoted here 
as B:Brookhaven; C:CRIEPI; M:MIT; T:Toronto; U:Univ. of Tokyo.}
\footnotetext[2]{Since the polarization direction was in the scattering 
plane in this experiment, it changed as momentum transfer was varied.}
\footnotetext[3]{The RIXS data were taken at $T=15$ K.}
\footnotetext[4]{SrCuO$_2$ has three different copper-oxygen bond lengths, 
which are represented as large error bars in Fig. 2.}
\end{table*}

\section{experiments}

The RIXS experiments were carried out at the Advanced Photon Source on the
undulator beamline 9IDB.  Experimental details have been described
elsewhere.\cite{LCO-PRL} Single crystal samples used in our measurements are
listed in Table~\ref{table1}, along with several samples studied in earlier
RIXS experiments. In Table~\ref{table1}, $d_{\rm Cu-O}$ and the experimental
configuration is listed for each material. All measurements were performed at
room temperature except for those on the LSCO17 sample. In our RIXS
experiments, the scattering plane was vertical and the polarization of the
incident x-ray, $\boldsymbol{\epsilon}$, was perpendicular to the scattering
plane. The polarization direction was kept fixed along the direction
specified in Table~\ref{table1}, where the coordinate system reference is the
$d_{x^2-y^2}$ orbital. That is, the copper oxygen plaquette lies in the
$xy$-plane, while the $z$-direction is perpendicular to the plaquette. We use
the notation of reduced momentum transfer {\bf q} throughout this paper, with
the $(\pi \; 0)$ direction along the Cu-O bond direction.

Before discussing the experimental results in detail, it is useful to first
review the second-order RIXS process to understand the nature of the observed
excitation. In the ground state, the holes are located in the antibonding
molecular orbital which is a combination of a Cu hole state ($d^9$) and an
oxygen ligand hole state ($d^{10}\underline{L}$), with more weight on the
$d^9$ state. In the intermediate state of this resonance process, a Cu $1s$
electron is excited to the Cu $4p$ band, and the core hole potential alters
the balance between the $d^9$ and the $d^{10}\underline{L}$ states. Then the
lowest energy state is predominantly $\underline{1s}d^{10}\underline{L}4p$,
which is lower than $\underline{1s}d^{9}4p$. These states form the so-called
well-screened and poorly-screened features, respectively, of the Cu K-edge
x-ray absorption spectra (XAS).  As discussed in detail by Hill and
coworkers, \cite{Hill98,Ide99} these intermediate states can decay into an
excited state in which the hole in the antibonding molecular orbital is
filled with an electron, creating a hole in the bonding orbital, and an
energy loss in the outgoing photon.  The RIXS process thus creates a
charge-transfer excitation from bonding to antibonding molecular orbitals.

In our measurements, we have carefully studied the incident energy
dependence in order to determine the resonance energy, for which the MO
excitation has the maximum intensity. In most cases, the resonance
energy, which is listed in Table~\ref{table1}, corresponds to the
higher energy peak (poorly-screened feature) in the XAS. On the other
hand, the intensity of lower energy excitations near the charge
transfer gap ($\sim 2$ eV) is strongly enhanced when the incident
energy corresponds to the well-screened intermediate state, as reported
in earlier studies.\cite{LCO-PRL} The resonance energy depends
on the direction of the polarization vector. Detailed
results of the incident energy dependence study will be published
elsewhere. However, one should note that the energy-loss associated
with the MO excitation does not depend on the incident polarization of
the photon, although different $4p$ states (e.g., $4p_{\sigma}$ or
$4p_{\pi}$) are involved in the intermediate state, as the polarization
is varied with respect to the $xy$-plane.  \cite{Hamalainen00}

\section{results}

\begin{figure}
\begin{center}
\epsfig{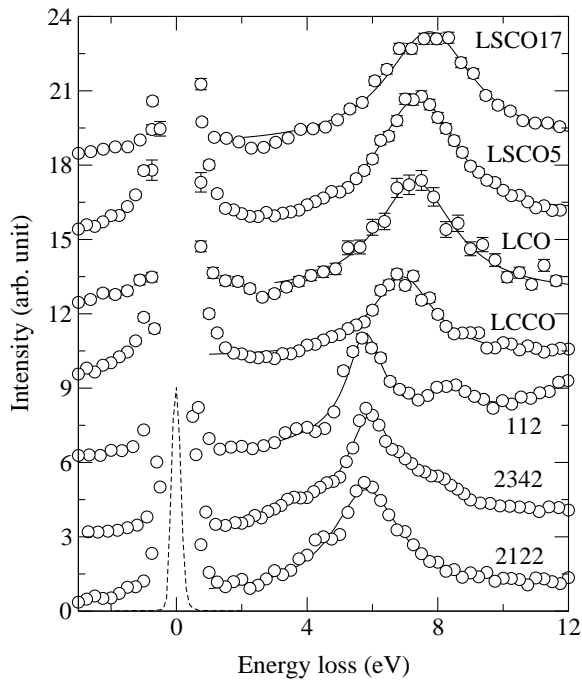}
\end{center}
\caption{RIXS spectra taken with the incident energy 
as specified in Table~\ref{table1} for a fixed reduced wave vector of ($\pi$ 0).
Each spectrum is offset vertically for clarity, and solid
lines are fits to a Lorentzian lineshape as described in the text.
The dashed line is a representative scan through the elastic line, 
which shows instrumental energy resolution.}
\label{fig1}
\end{figure}

\begin{figure}
\begin{center}
\epsfig{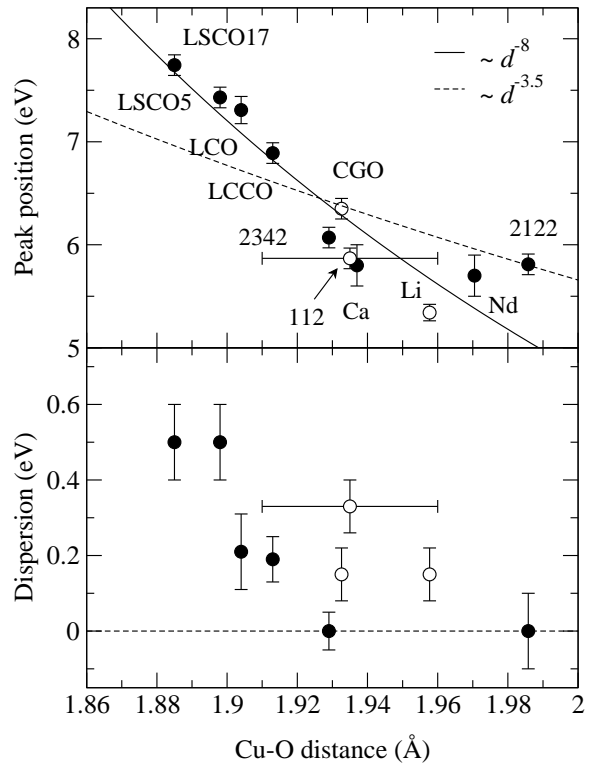}
\end{center}
\caption{(a) 
The value of the peak position at ($\pi$ 0), $\Delta_{\sigma\sigma^\ast}$, and
(b)  the amount of dispersion along the Cu-O bond direction for each sample is 
plotted as a function of
$d_{\rm Cu-O}$. The solid symbols are used for the perfect square plaquettes,
while the open symbols are for the samples with distorted plaquettes. The solid 
and dashed lines are fits to
power law expressions $\Delta_{\sigma\sigma^\ast} \sim d_{\rm Cu-O}^{-8}$ and 
$d_{\rm 
Cu-O}^{-3.5}$, respectively. Note that the 
measured dispersion of the edge-sharing chain 
compounds Li and CGO is not along the Cu-O bond direction, while dispersion was 
not measured for Nd and Ca.\cite{Hill98,Hasan00}}
\label{fig2}
\end{figure}

In Fig.~\ref{fig1}, representative RIXS scans are plotted.  These are
energy-loss scans taken at a fixed momentum transfer with the incident
energy of the x-ray photon fixed at the values listed in
Table~\ref{table1}. The momentum transfer for all these scans has been
fixed at {\bf q}=($\pi$ 0), which is the minimum energy position. The most
striking feature in Fig.~\ref{fig1} is the large shift of the excitation
energy from $\sim 6$ eV for 2122 to $\sim 8$ eV for LSCO17. To analyze this
shift quantitatively, we have fitted the observed excitation spectra to a
simple Lorentzian lineshape and extracted the peak positions, which are
plotted in Fig.~\ref{fig2}(a) as a function of Cu-O bond length, 
$d_{\rm Cu-O}$.

\begin{figure}
\begin{center}
\epsfig{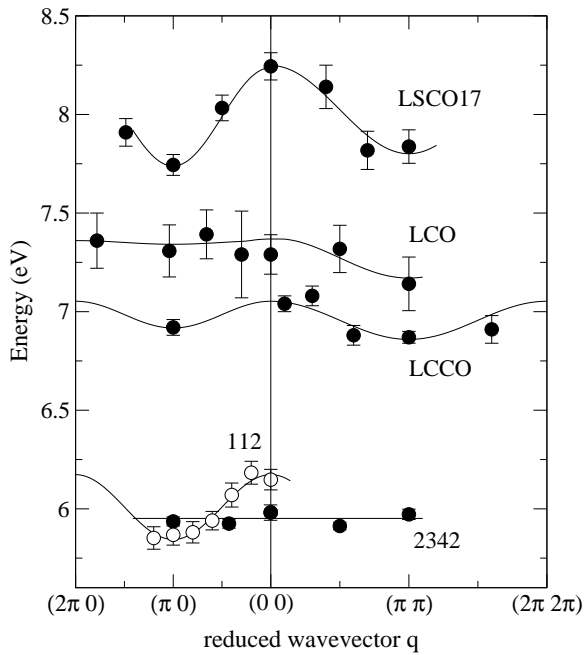}
\end{center}
\caption{The observed dispersion of the MO excitation.
The left panel is along the ($\pi$ 0) 
direction, and 
the right panel is data taken along the ($\pi$ $\pi$) direction.
The solid lines are guides to the eye.} 
\label{fig3} 
\end{figure}

A few comments are in order regarding the data analysis.  First, as is
evident from the instrumental resolution plotted as a dashed line in
Fig.~\ref{fig1}, the observed excitations are not resolution-limited, hence
justifying our simple fitting procedure, which does not convolve the data
with the instrumental resolution. Second, in several cases, we observe more
than one type of excitation in these scans. For example, the SrCuO$_2$ data
clearly shows a second feature around $\sim 8.2$~eV, in addition to the main
peak around $\sim 6$~eV. In the 2342 case, there also seems to be two
additional features, one at higher energy ($\sim 8$ eV) and the other at
lower energy ($\sim 4$ eV). However, it is difficult to identify these weak
features (if present) in the data for the other samples,\cite{Abbamonte} and
we have chosen to fit all the scans with a single Lorentzian peak with a
broad width. The peak positions extracted from our analysis are, therefore,
those of the dominant features. In addition, the peak width extracted from
the fits might in some cases arise from a distribution of several peaks over
a wide energy range, rather than from a finite inverse lifetime of a single
excitation.  Finally, as discussed below, we observe dispersion of the MO
excitation with momentum transfer. This can be as large as $\sim 0.5$ eV in
some of the compounds studied, as shown in Fig.~\ref{fig3}. Thus, the peak
position of the MO excitation depends not only on the sample, but also on
{\bf q}. The scans shown in Fig.~\ref{fig1} all are taken at the minimum
energy position, ${\bf q}=(\pi \; 0)$, and the peak positions plotted in
Fig.~\ref{fig2}(a) are the values measured at this position.

As shown in Fig.~\ref{fig2}(a), the energy of the MO excitation exhibits a
strong dependence on $d_{\rm Cu-O}$. It is noteworthy that the excitation
energy exhibits such a systematic dependence on the {\it local} structure,
and that it is apparently insensitive to whether the crystal has a planar,
corner-sharing, or edge-sharing chain strucutre. This is consistent with our
assignment of these features as MO excitations localized within a single  
Cu-O plaquette.

The overall trend exhibited in Fig.~\ref{fig2}(a) is not unexpected.
Intuitively, as the Cu and O atoms move closer, the $p-d$ overlap will
increase, and the Cu-O bonding becomes more covalent with a larger energy
splitting $\Delta_{\sigma\sigma^\ast}$. What is surprising is how strong this
$d_{\rm Cu-O}$ dependence is. We have modeled the $d_{\rm Cu-O}$-dependence
of $\Delta_{\sigma\sigma^\ast}$ as a power law ($\Delta_{\sigma\sigma^\ast}
\sim d^\eta$) and find $\eta=-8(2)$, shown as a solid line in
Fig.~\ref{fig2}(a).  Note that the MO excitation energy is expected to be
given by\cite{vanVeenendaal93}
$\Delta_{\sigma\sigma^\ast}=\sqrt{(2t_{pp}-\Delta_0)^2+16t_{pd}^2}$, where
$t_{pp} \approx 0.65$ eV is the hopping matrix element between the oxygen $p$
orbitals.\cite{Hybersten89} Since $16t_{pd}^2 \gg (2t_{pp}-\Delta_0)^2$, this
expression leads to $\Delta_{\sigma\sigma^\ast} \approx 4t_{pd}$, to a first
approximation, and one then expects similar a $d$-dependence for
$\Delta_{\sigma\sigma^\ast}$ and $t_{pd}$.  Our results then implies that the
$d$-dependence of $t_{pd}$ is much stronger than that expected from
tight-binding theory, for which $\sim d^{-3.5}$ is
predicted.\cite{Harrison89} As plotted in Fig.~\ref{fig2}(a), 
the observed RIXS data clearly deviates from the $\Delta_{\sigma\sigma^\ast} 
\sim d^{-3.5}$ behavior (dashed line). 
We also note that the $t_{pd} \sim d^{-8}$
behavior determined from our RIXS measurements is different from the earlier
report of $t_{pd} \sim d^{-4}$ by Cooper and coworkers, which was estimated
indirectly from a three-band Hubbard model expression.\cite{Cooper90}

In Fig.~\ref{fig3}, the dispersion of the MO excitation is plotted as a
function of {\bf q} for selected samples. These data suggest that the
picture of a completely localized MO excitation is an
oversimplification -- such a model would predict no
dispersion of these features. The data of Fig.~\ref{fig2}(b), which show
significant dispersion for $d_{\rm Cu-O} < 1.93$ \AA, suggest then that this
localized picture breaks down for small bond distances. For the LSCO17
sample, the dispersion bandwidth is about $0.5$ eV, with the minimum
excitation energy occuring at the zone boundary, implying an indirect
gap.  Note that this is completely different from the direct nature of
the lower energy charge-transfer gap as reported in 
Ref.~\onlinecite{LCO-PRL}.  As
$\Delta_{\sigma\sigma^\ast}$ decreases (e.g., LCO and LCCO), the
dispersion of the MO excitation becomes weaker. For the samples with an
even smaller $\Delta_{\sigma\sigma^\ast}$, the dispersion becomes flat,
as shown for the 2342 sample (Figs. 2 and 3). Limited momentum
dependence measurements for the 2122 sample (not shown) also show a
dispersionless behavior. The size of the observed dispersion along the
Cu-O bond direction is plotted against $d_{\rm Cu-O}$ in
Fig.~\ref{fig2}(b).

As shown in Fig.~\ref{fig2}, the size of the dispersion and  
$\Delta_{\sigma\sigma^\ast}$ both increase as $d_{\rm
Cu-O}$ decreases. This suggests that the bandwidth of the dispersion is
also controlled by the hopping parameter $t_{pd}$. Such behavior would, of
course, be expected from the increased overlap of the wavefunctions, since
charge carriers then become less localized. However, the simplest picture
of delocalized electrons fails to describe the observed dispersion. For
example, the bandwidths of the Cu-O bonding and antibonding bands in LCO
are very large $\sim 3$ eV, and interband transitions between these two
bands would have a {\it direct} gap of $\sim 4$ eV.
\cite{Mattheiss87,McMahan88}

\section{discussion and summary}

One of the most surprising results in this study is that the
$\Delta_{\sigma\sigma^\ast}-d_{\rm Cu-O}$ scaling seems to apply to
different structures. In contrast, previous studies of the 
bond-length
scaling of various quantities, such as charge-transfer gap 
($\Delta_{CT}$),
or superexchange interaction ($J$), have been limited to compounds with
corner-sharing structures.\cite{Ohta91} It is well known that such
quantities as superexchange coupling depend not only on the $p-d$
hybridization, but also crucially on the angle between the two Cu-O
bonds.\cite{Mizuno98} One can argue that
$\Delta_{\sigma\sigma^\ast}$ is a better measure of Cu-O hybridization
than $J$ or $\Delta_{\rm CT}$, since it is only dependent on $d_{\rm 
Cu-O}$ and not on the presence of neighboring 
atoms.

We have noted that the $t_{pd} \sim d^{-8}$ dependence inferred from our
study deviates significantly from the tight-binding picture.  It also appears
to give rise to discrepancies with other experiments. For example, for
materials with the corner-sharing structure, the three-band Hubbard model
gives $J \sim t_{pd}^4/U^\ast\Delta_{CT}^2$, where $U^\ast$ is an effective
onsite Colomb interaction which is assumed to be constant.  If we use the
experimentally determined\cite{Cooper90} $\Delta_{CT} \sim d^{-6}$ and our
$t_{pd} \sim d^{-8}$ result, we obtain $J \sim d^{-20}$, which clearly
disagrees with the much weaker $d$-dependence of $J$ observed in various
experiments, including two magnon Raman scattering.\cite{Sulewski90} One
might expect that this discrepancy could be resolved by considering the fact
that, due to strong electron correlations, the simple tight-binding picture
of covalent bonding has to be modified. In fact, Mizuno and
coworkers\cite{Mizuno98r} considered two contributions to $t_{pd}$. That is,
$t_{pd}=t_{pd}^0+t_{pd}^M$, where $t_{pd}^0$ is the contribution from the
atomic potential which depends only on $d_{\rm Cu-O}$, while $t_{pd}^M$ is
the contribution from the Madelung potential, which depends on the detailed
arrangement of the neighboring ions. However, the calculated contribution
from the Madelung potential is of order of $\sim 0.1$ eV or
smaller,\cite{Mizuno98r} so that this alone is not enough to explain the
$\sim d^{-8}$ dependence.

These results may be suggesting that one has to abandon the simple
relationship of $\Delta_{\sigma\sigma^\ast} \approx 4t_{pd}$. Certainly, as
discussed above, the picture of a completely localized MO excitation is
apparently an oversimplification, since it breaks down as $d_{\rm Cu-O}$
becomes shorter -- as evidenced by the sizable dispersion observed in LSCO17.
Thus, if a more realistic expression for $\Delta_{\sigma\sigma^\ast}$ is
used, $t_{pd} \sim d^{-3.5}$ scaling law might be recovered.  
For example, a recent first-principles calculation has emphasized the role of
apical oxygens in the systematics of high temperature superconductivity
\cite{Pavarini01}.  Indeed, the scaling plot in Fig.~\ref{fig2} also exhibits
some systematic dependence on the number of apical oxygens,\cite{a_O} and it
may be interesting to further investigate the role played by apical oxygens.
Certainly, a systematic {\it ab initio} calculation of the $d_{\rm
Cu-O}$-dependence of MO excitation energy in large clusters would be highly
desirable and may help to clarify the relationship between
$\Delta_{\sigma\sigma^\ast}$, $t_{pd}$, and $d_{\rm Cu-O}$.

To summarize, we have studied a charge-transfer excitation in various cuprate
compounds using resonant inelastic x-ray scattering technique. We assign this
excitation to a mostly localized molecular orbital excitation, that is, an
excitation from a bonding to an antibonding molecular orbital.  We have found
that this molecular orbital excitation energy, which is a measure of the
hopping matrix element $t_{pd}$, exhibits a systematic Cu-O bond length
dependence, which is much stronger than that expected from tight-binding
theory. We have also observed a sizable dispersion of this excitation in some
materials, suggesting that this molecular orbital excitation becomes less
localized as the $p-d$ hybridization becomes large.

\acknowledgements{
We would like to thank P. D. Johnson, G. A. Sawatzky, and M. A. van 
Veenendaal for invaluable
discussions. The work at Brookhaven was supported by the U. S. Department of
Energy, Division of Materials Science, under contract No. DE-AC02-98CH10886. Use
of the Advanced Photon Source was supported by the U. S. Department of Energy,
Basic Energy Sciences, Office of Science, under Contract No.  
W-31-109-Eng-38.}


\begin{thebibliography}{10}
\bibitem{a_O}
There are two apical oxygens in the structure of LSCO17, LSCO5, and LCO, 
while there is one apical oxygen for LCCO. The rest of the samples have 
no apical oxygens.
\end{thebibliography}
\end{document}